\documentclass{PoS}

\newcommand{\sign}{{\rm sign}}
\newcommand{\les}{\stackrel{<}{{}_{\sim}}}

\title{Exploring the chiral regime with dynamical overlap fermions}

\ShortTitle{Exploring the chiral regime with dynamical overlap fermions}

\author{\speaker{Hideo Matsufuru}%
  \ for the JLQCD%
  \thanks{JLQCD Collaboration:
       S.~Aoki, H.~Fukaya, S.~Hashimoto, K-I.~Ishikawa, K.~Kanaya,
       T.~Kaneko, H.~M., J.~Noaki, E.~Shintani, M.~Okawa, T.~Onogi, 
       A.~Ukawa, N.~Yamada, T.~Yoshi\'e.}
  \ and TWQCD%
  \thanks{TWQCD Collaboration:
       T-W.~Chiu, T-H.Hsieh, K.~Ogawa.}
  \  Collaborations\\
        High Energy Accelerator Research Organization (KEK),
        Tsukuba 305-0801, Japan \\
        E-mail: \email{hideo.matsufuru@kek.jp}}

\abstract{
I report the status of the dynamical overlap fermion project by the
JLQCD and TWQCD collaborations.
So far, the simulations have been completed with two flavors of overlap
sea quarks in a wide range of sea quark mass corresponding to
the $p$-regime and
the $\epsilon$-regime on a $16^3\times 32$ lattice at $a=$0.12 fm.
More recently, runs with 2+1 flavors of sea quarks have also started.
This talk mainly discusses the physics results on the $N_f=2$ lattice
after describing the lattice formulation
and algorithms.}

\FullConference{The XXV International Symposium on Lattice Field Theory\\
		 July 30 - August 4 2007\\
		 Regensburg, Germany}

\begin{document}

\section{Introduction}

Recent development of algorithms and computational resources has
enabled extensive studies of dynamical lattice QCD simulations
at small quark masses where the chiral symmetry plays an essential
role.
Such studies are indispensable to investigate the chiral dynamics
of QCD and to determine hadronic matrix elements with the precision
required by present and future flavor physics.
Among lattice fermion actions, the overlap fermion
\cite{Neuberger:1997fp,Neuberger:1998wv}
has attractive features for these purposes.
The overlap fermion operator is represented as
\begin{equation}
D = m_0 \left[ 1+ \gamma_5 \, \sign(H_W(-m_0)) \right],
\label{eq:overlap_opr}
\end{equation}
where $H_W=\gamma_5 D_W$ is the hermitian Wilson-Dirac operator
with a large negative mass $-m_0$.
This operator satisfies the Ginsparg-Wilson relation
\cite{Ginsparg:1981bj},
and thus holds an exact chiral symmetry on the lattice
\cite{Hasenfratz:1998ri,Luscher:1998pq}.
The exact chiral symmetry significantly simplifies the structure of
operator mixing in calculations of the matrix elements.
The overlap fermion corresponds to the infinite $N_s$ limit of
the domain-wall fermion, which means that one does not have to
take care of the residual mass.
On the other hand, numerical implementation of the overlap fermion
is expensive,
because of the evaluation of the sign function of $H_W$.
Furthermore, the discontinuity of the operator at $\lambda=0$,
where $\lambda$ is an eigenvalue of $H_W$, makes the
molecular-dynamics evolution much involved.
Therefore, dynamical simulations of the overlap fermion have become
feasible only recently with improved algorithms and large
computational power.

We are running a large-scale lattice QCD simulation project 
with 2 and 2+1 flavors of dynamical overlap fermions.
Our physics goals are to explore the chiral regime of QCD with
an exact chiral symmetry, and to compute hadronic matrix elements
with controlled chiral extrapolation.
Present simulations are performed with a spatial lattice size
of 16 and $a\simeq$ 0.12 fm.
For the gauge action, we adopt the Iwasaki's renormalization group
improved action, as well as a topology fixing term which suppresses
near-zero modes of $H_W$
\cite{Vranas:1999rz,Fukaya:2006ca,Fukaya:2006vs}.
Thus our simulation is performed in a fixed topological charge sector.
By avoiding the discontinuity of the overlap operator at $\lambda=0$,
the numerical cost of HMC is significantly reduced.

This report explains our strategy of simulations and presents
several recent results.
In the next section, we address why the fixed topology simulation
is feasible, and how it can extract physical observables with
controlled systematic errors.
Section~\ref{sec:algorithm} describes the numerical algorithms
which are essential for dynamical overlap simulations.
In Section~\ref{sec:simulation}, our simulation set-up is
summarized.
Some of recent results at $N_f=2$ are shown in
Section~\ref{sec:results}.
The last section is devoted to a conclusion and outlook.

\section{Simulations at fixed topology}

\subsection{Topology fixing term}

Dynamical simulations of the overlap fermions are quite nontrivial,
because the sign function in Eq.~(\ref{eq:overlap_opr}) has
a discontinuity at $\lambda=0$, where $\lambda$ is an eigenvalue of $H_W$.
Vanishing $\lambda$ may occur during molecular-dynamics steps of
the HMC update, at which the topological charge of the system changes
its value.
To keep the acceptance rate of HMC, one needs to take care of
this discontinuity so as to conserve the Hamiltonian precisely.
One option is the reflection/refraction prescription
\cite{Fodor:2003bh}.
This method first determines the time step at which $\lambda$
vanishes, and there the change of the pseudofermion action is
computed.
Like the light traveling across a surface of water,
if the momentum of the mode is larger than the change of
the pseudofermion action, it is refracted, while otherwise
reflected.
This method requires additional inversions of $D(m)$ at $\lambda=0$,
and thus the numerical cost quickly increases when the low-mode density
becomes large.

We instead employ a topology fixing term, which is implemented
with an extra Wilson fermion and a twisted mass ghost as
\cite{Vranas:1999rz,Fukaya:2006ca,Fukaya:2006vs}
\begin{equation}
  \det\left( \frac{H_W^2}{H_W^2+\mu^2} \right)
 = \int {\cal D}\chi^\dag  {\cal D}\chi \exp(-S_E),
\label{eq:extra-Wilson}
\end{equation}
\begin{equation}
  S_E = \chi^\dag \left[ (D_W+i\gamma_5\mu)(D_W^\dag D_W)^{-1}
  (D_W + i\gamma_5 \mu)^\dag \right] \chi.
\label{eq:extra-Wilson2}
\end{equation}
This term is irrelevant in the continuum limit,
and considered as a part of the gauge action.
The numerator of Eq.~(\ref{eq:extra-Wilson}) suppresses
near-zero modes of $H_W$, while contribution from large
frequency modes are compensated by the denominator.
Since $\lambda=0$ is prohibited, the topological charge
is fixed during the molecular-dynamics update, and hence
the reflection/refraction is no longer needed.
We set the twisted ghost mass $\mu=0.2$ throughout this work.
As was shown in Refs.~\cite{Fukaya:2006vs,Kaneko:2006pa,
Hashimoto:2006rb}, this term successfully suppresses the
near-zero modes of $H_W$.

As an alternative approach, the tunneling HMC
has been proposed recently in Ref.~\cite{Golterman:2007ni}.
This method also employs the extra Wilson fermion term,
and thus near-zero modes of $H_W$ are suppressed,
but it projects out a few lowest-lying modes during the
molecular-dynamics steps so as to enable topology changes.
The tunneling HMC does not avoid $\lambda=0$, and the
reflection/refraction prescription is necessary.
Practical feasibility test is to be performed.

\subsection{Simulations at fixed topology}
 \label{subsec:Simulation_fixed_topology}

Simulations at fixed topology are especially useful in the
$\epsilon$-regime where $1/m_\pi \gg L$
(see Sec.~\ref{subsec:epsilon-regime}).
In the $\epsilon$-regime, the topological charge dependence of
physical observables is manifest and the fixed topology simulation
is useful to determine the low-energy constants appearing in the
effective chiral Lagrangian.
On the other hand, in the ordinary regime ($p$-regime), precision
calculations are possible only when the following two conditions
are satisfied;
(1) A physical observable in the $\theta$ vacuum is related to
those in the fixed-$Q$ vacua as a systematic expansion in terms of
$V^{-1}$, where $V$ is the spacetime volume.
(2) The topological susceptibility, $\chi_t=\langle Q^2 \rangle /V$,
is calculable and reproduces the known behavior from ChPT,
which ensures that
the local fluctuation of topological charge is active
so as to produce relevant physics in a finite volume.
Here we explain these points along
Refs.~\cite{Brower:2003yx,Aoki:2007ka,Talk_Onogi}.

What we want to compute is an expectation value in the $\theta$
vacuum, which is related to the ``vacua'' with fixed values of $Q$
through the Fourier transformation,
\begin{equation}
 Z(\theta) = \sum_{Q} e^{-i\theta Q} Z_Q,
\hspace{0.5cm}
 Z_Q = \int_{-\pi}^{\pi} \frac{d\theta}{2\pi} e^{i\theta Q} Z(\theta).
\end{equation}
While the cluster property does hold for the former,
the global topology becomes irrelevant in the infinite volume limit
and the local fluctuation of topological charge is responsible to
the physics.
In practice, however, the volume is inevitably finite.
For a large enough volume, $\chi_t V\gg 1$, and for $Q \ll \chi_t V$,
the saddle point analysis is applicable and it leads to the conclusion
that the distribution of $Q$ is Gaussian and physical observables are
represented as $\langle O \rangle_Q = \langle O \rangle_\theta
 + \mbox{(finite V corrections)}$.
As an explicit example, a CP even correlator is represented as
\begin{equation}
 G_Q = G(0)
   + G^{(2)}(0) \frac{1}{\chi_t V}
      \left[ 1 - \frac{Q^2}{\chi_t V} - \frac{c_4}{2\chi_t^2 V}
        \right]
   + G^{(4)}(0) \frac{1}{8\chi_t^2 V^2} + O(V^{-3}),
\label{eq:Corr_fixedQ}
\end{equation}
where $G^{(n)}(\theta)$ is the $n$-th derivative
of $G(\theta)$ with respect to $\theta$, and
$c_4=-(\langle Q^4 \rangle - 3\langle Q^2 \rangle^2)/V$.
To quantify the $O(1/V)$ effect in Eq.~(\ref{eq:Corr_fixedQ}),
$G^{(2)}$ must be known.
This is determined with a help of the chiral perturbation theory (ChPT).
For example, the $O(V^{-1})$ fixed topology effect to the PS meson
mass is calculated in Ref.~\cite{Brower:2003yx} at the tree-level of
ChPT.
It is also possible to determine $G^{(2)}$ in numerical simulations
by comparing the results in different $Q$ sectors.

In the above formula, the topological susceptibility $\chi_t$ plays
a key role.
For a self-contained calculation, $\chi_t$ must be computed on the
fixed-$Q$ configurations.
$\chi_t$ is represented as a correlation of the local topological
charge $\omega(x)$, which is also subject to Eq.~(\ref{eq:Corr_fixedQ}).
From the clustering property in the $\theta$ vacuum,
\begin{equation}
 \lim_{|x|\rightarrow \infty} \langle \omega(x) \omega(0) \rangle_Q
   = \frac{1}{V}\left( \frac{Q^2}{V} - \chi_t
                           - \frac{c_4}{2\chi_t V} \right)
   +  {\cal O}(V^{-3}).
\end{equation}
Using the axial Ward-Takahashi identity,
\begin{equation}
 \lim_{|x|\rightarrow \infty} \langle mP^0(x) mP^0(0) \rangle_Q
 = \lim_{|x|\rightarrow \infty} \langle \omega(x) \omega(0) \rangle_Q ,
\end{equation}
where $P^0(x)$ is the flavor singlet pseudoscalar density.
Thus the topological susceptibility can be extracted from the
long range behavior of the correlation function carrying
the quantum number of $\eta'$ meson.
If the determined $\chi_t$ in a simulation exhibits a reasonable value,
it implies that in that system the local fluctuation of topological
charge is active enough, and the system size is sufficiently large.
Extraction of $\chi_t$ in our $N_f=2$ simulation will be described
in Sec.~\ref{subsec:topological_susceptibility}.
Such a method is useful not only in the fixed topology simulations
but also in standard HMC updates, because the changes of topological
charge with a continuous variation of link variables
become increasingly rare as approaching the continuum limit.

These arguments indicate that the fixed topology simulations can
provide a framework to determine the physical observables in the
$\theta$ vacuum in a self-contained manner with the finite size
effects under control.

\section{Algorithm}
\label{sec:algorithm}

\subsection{Overlap operator}

The overlap operator with a quark mass $m$ is written as
\begin{equation}
 D(m) = \left( m_0 + \frac{m}{2} \right)
 +   \left( m_0 - \frac{m}{2} \right) \gamma_5 \sign[H_W(-m_0)].
\end{equation}
$m_0$ is set to 1.6 throughout this work.
The sign function means that an eigenmode
$(\lambda, v_\lambda)$ of $H_W$ is transformed to
$(\sign(\lambda),v_\lambda)$.
Since the calculation of all the eigenmodes is impractical,
some kind of approximation of the sign function is required.
We employ the Zolotarev rational approximation
\cite{vandenEshof:2002ms,Chiu:2002eh},
\begin{equation}
\frac{1}{\sqrt{H_W^2}} = \frac{d_0}{\lambda_{min}}
  (h_W^2 + c_{2n}) \sum^N_{l=1} \frac{b_l}{h_W^2 + c_{2l-1}} ,
\label{eq:Zolotarev}
\end{equation}
where $h_W=H_W/\lambda_{min}$ with $\lambda_{min}$ the
eigenvalue having the smallest absolute value.
$d_0$, $c_{l}$, $b_l$ are easily calculable parameters.
Although $(h_W^2 + c_{2l-1})^{-1}$ must be calculated $N$ times,
these terms can be obtained simultaneously by the multi-shift CG
method \cite{Frommer:1995ik,Jegerlehner:1996pm}.
Thus, the numerical cost mildly depends on $N$.
This formula is valid in the region
$|\lambda | \in [\lambda_{min},\lambda_{max}]$.
Since the error of the formula scales as $\exp(-\lambda_{min}N)$,
the smaller $\lambda_{min}$ requires the larger $N$ to keep the
precision unchanged.
Instead, if one calculates low-lying eigenvalues of
$H_W$ of $|\lambda|<\lambda_{thrs}$,
one can determine the sign function of these modes explicitly and
project them out from $H_W$.
Then $\lambda_{thrs}$ replaces $\lambda_{min}$ in the above formula,
leading to
\begin{equation}
 \sign (H_W)  = \sum_{j=1}^{N_{ev}} \sign(\lambda_j) v_j \otimes v_j^\dag
       + \sign (H_W) P_H,
\end{equation}
where $P_H=1-\sum_{j=1}^{N_{ev}} v_j \otimes v_j^\dag$, and $N_{ev}$
the number of modes with $|\lambda_j|<\lambda_{thrs}$.
The approximation formula is applied to the second term of this equation.
The numerical cost depends on the density of low modes, since the
determination of the low modes requires non-negligible computational
time, and also projecting out the low modes from the sign function
requires additional operations.
Also in this sense, employing the topology fixing term improves
the  simulations of overlap fermions.
In this work, we adopt $\lambda_{thrs}=0.045$ and $N=10$,
which lead to an accuracy of $|\sign^2 H_W -1|\simeq 10^{-(7-8)}$.

\subsection{Solver algorithms}

Since the overlap operator must be inverted at each step of the
molecular-dynamics evolution, the improvement of the solver algorithm
may significantly reduce the simulation cost.
We have tested two algorithms; the nested CG (or 4DCG)
method \cite{Cundy:2004pz} and the 5-dimensional CG (5DCG) method
\cite{Borici:2004pn,Edwards:2005an}.
The nested CG method is a straightforward implementation of the
overlap solver.
It contains two nested CG iterations: an outer loop for operating $D(m)$
and an inner loop for the calculation of $(H_W^2 + c_{2l})^{-1}$.
The numerical cost can be reduced by applying the relaxation technique,
which relaxes the convergence criterion of the inner loop as the outer
loop iteration proceeds.
Instead of the CG method, one can apply other Krylov subspace algorithms
such as GMRES and SUMR.
Since in HMC only the inversion of $D^\dag D$ appears and the CG method
almost achieves the best performance, we compare the CG method with
the 5D algorithm in the following.

The 5-dimensional CG solver is based on the Schur decomposition
\cite{Borici:2004pn,Edwards:2005an}.
Let us consider a 5-dimensional block matrix
(the $N=2$ case is displayed as an example),
\begin{equation}
 M_5 =
  \left( \begin{array}{cccc|c}
               H_W        & -\sqrt{q_2} & & & 0  \\
            -\sqrt{q_2} & -H_W          & & & \sqrt{p_2} \\
               & &    H_W        & -\sqrt{q_1} & 0  \\
               & & -\sqrt{q_1} & -H_W          & \sqrt{p_1} \\
    \hline
             0 &\sqrt{p_2} & 0 & \sqrt{p_1} & R \gamma_5 + p_0 H \\
         \end{array}   \right)
 = 
  \left( \begin{array}{c|c}
               A   & B  \\
    \hline
               C   & D  \\
         \end{array}   \right) .
\label{eq:5Dmatrix}
\end{equation}
Since $M_5$ can be decomposed as
\begin{equation}
 M_5 =
  \left( \begin{array}{cc}
             1 & 0 \\
             CA^{-1} & 1 \\
         \end{array}   \right)
  \left( \begin{array}{cc}
             A & 0 \\
             0 & S \\
         \end{array}   \right)
  \left( \begin{array}{cc}
             1 & A^{-1}B \\
             0 & 1 \\
         \end{array}   \right) ,
\end{equation}
where $S=D-CA^{-1}B$ is called the Schur complement.
One can solve a 4D linear equation $S \psi_4 = \chi_4$ by
solving a 5D equation
\begin{equation}
 M_5
  \left( \begin{array}{c}
             \phi   \\
             \psi_4 \\
         \end{array}   \right)
= \left( \begin{array}{c}
               0   \\
             \chi_4 \\
         \end{array}   \right) .
\label{eq:5dim_lineq}
\end{equation}
Setting the parameters $R$, $p_0$, $p_i$ and $q_i$ ($i=1,\dots,N$)
in Eq.~(\ref{eq:5Dmatrix}) appropriately,
the 5D solver can be used to invert the overlap operator
approximated by Eq.~(\ref{eq:Zolotarev}).
The 5D solver is accelerated by applying the even-odd
preconditioning.
Then one needs to solve a reduced linear equation,
$(1-M_{ee}^{-1}M_{eo}M_{oo}^{-1}M_{oe})\psi_e=\chi'_e$, where
even/odd blocks of $M_5$ is denoted by $M_{ee}$, $M_{eo}$, etc.
The inversions $M_{ee}^{-1}$ and $M_{oo}^{-1}$ are easily calculated
by forward/backward substitutions.
The projection of low-modes is not straightforward for the even-odd
preconditioned 5D solver, since the operation of $M_{ee}^{-1}$
(or $M_{oo}^{-1}$) becomes much involved.
Nevertheless it can be implemented cheaply, because the subspace of
the matrix is spanned by $x_e$, $\gamma_5 x_e$,
$v_{je}$, $\gamma_5 v_{je}$ ($j=1,\dots ,N_{ev}$)
\cite{Poster_Hashimoto}.

\begin{figure}[tb]
\center{
\includegraphics[clip=true,width=8.0cm]{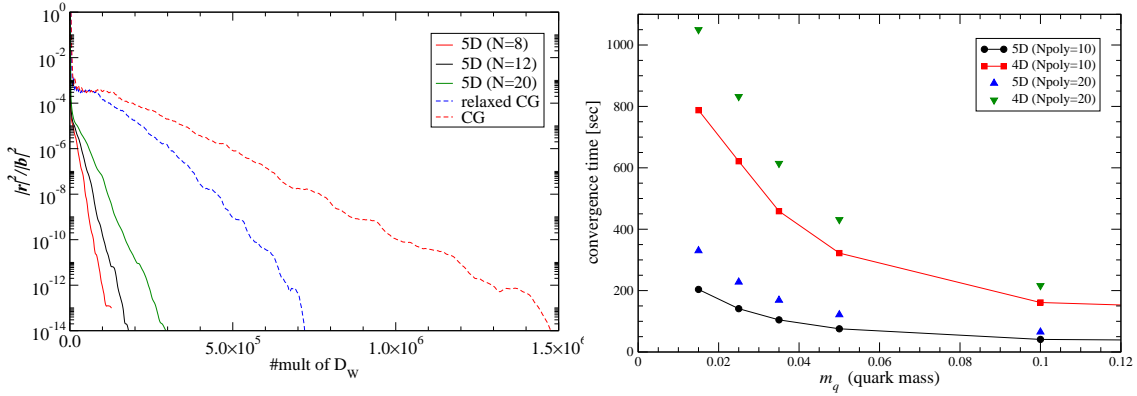}
\hspace{-0.5cm}
\includegraphics[clip=true,width=7.3cm]{Figs/conv_vs_mq.eps}
}
\vspace{-0.6cm}
\caption{
Comparisons of the overlap solvers.
The left panel shows the residual of the CG solver without
the low-mode projection.
The right panel shows the required time for the convergence
in the case with the projection of 8 low modes of $H_W$.
}
\label{fig:solver}
\vspace{-0mm}
\end{figure}

Figure~\ref{fig:solver} compares the 4D and 5D solvers.
The left panel shows the behavior of the residual against
the numbers of the Wilson-Dirac operator multiplications,
in the case of no low-mode projection \cite{Matsufuru:2006xr}.
The relaxation accelerates the nested CG by a factor of two,
while the 5D solver
exhibits much faster convergence for practical values of $N$.
The required time to solve the linear equation is compared in
the right panel for the case that the number of projected low
modes is 8.
This shows that the 5D solver is 3--4 times faster than the 4D
solver in the whole region of quark mass used in this work.
Therefore we mainly use the 5D solver in HMC.
For the computation of the quark propagator, we adopt the nested
4D solver because of an advantage to obtain the propagators
with several valence quark masses simultaneously with the
multi-shift CG algorithm.

Here let us compare the cost of simulations with the domain-wall
fermion.
Let us consider the cost to solve a linear equation $Dx=b$.
At $m\simeq m_s/2$ , $a^{-1}\simeq 1.7$ GeV, and on $16^3\times 32$
lattices, the domain-wall solver
requires $O(800)$ iterations for a precision $|r|/|b|<10^{-10}$
according to Ref.~\cite{Aoki:2004ht},
while the overlap solver requires $O(1200)$ iterations.
To convert to the numbers of the Wilson-Dirac operator multiplications,
$2N_s$ and $2N+1$ are multiplied for the domain-wall and the overlap
fermions, respectively.
Then the difference amounts to a factor of about 2.5.
In HMC, the overlap solver must be called twice, while the domain-wall
solver is applied to $D^\dag D$.
This leads to another factor of 2 difference leading to the total
difference of $O(5)$.

\subsection{Hybrid Monte Carlo algorithm}

Implementation of the HMC algorithm for the overlap operator
with the approximation formula (\ref{eq:Zolotarev}) is
straightforward except for the discontinuity at $\lambda=0$.
Since we employ the topology fixing term, $\lambda =0$ does
not appear and no reflection/refraction prescription is required.
In order to improve the performance of HMC, we adopt
the mass preconditioning \cite{Hasenbusch:2001ne} together
with the multi-time step procedure \cite{Sexton:1992nu}.
Introducing a preconditioning term with a heavier quark
mass $m'$ than the dynamical quark mass $m$,
the fermion action becomes $S_F = S_{PF1} + S_{PF2}$,
\begin{equation}
 S_{PF1} = \phi_1^\dag [D(m')^\dag D(m')]^{-1} \phi_1,
 \hspace{0.3cm}
 S_{PF2} = \phi_2^\dag \left\{ D(m')
          [D(m)^\dag D(m)]^{-1}  D(m')^\dag \right\} \phi_2 .
\end{equation}
The forces from the preconditioner (PF1),
the preconditioned dynamical quark (PF2), the gauge field (G), and 
the extra Wilson fermion/ghost field (E) have a hierarchical structure,
\begin{equation}
 F_{G} \sim  F_{E} \gg  F_{PF1}  \gg F_{PF2} .
\end{equation}
Thus we set
\begin{equation}
 \Delta \tau_{(PF2)} \gg \Delta \tau_{(PF1)} \gg
 \Delta \tau_{(G)} = \Delta \tau_{(E)} .
\end{equation}
$\Delta \tau_{(E)}$ is set to be the same as the gauge part 
also to ensure the disappearance of near-zero modes of $H_W$.
The cost to calculate $F_E$ is negligible compared to the overlap
fermions.
In the case of $N_f=2+1$, the time steps of the two-flavor and
one-flavor parts are set equal.
In this work, $\Delta \tau_{(PF2)}/\Delta \tau_{(PF1)}$ and
$\Delta \tau_{(PF1)}/\Delta \tau_{(G)}$ are set to 4--6.

In the $N_f=2$ simulation, we employ the noisy Metropolis
test \cite{Kennedy:1985pg} together with a less precise
5D solver without the projection of low-modes of $H_W$
\cite{Kaneko:2006pa,Matsufuru:2006xr}.
During the molecular dynamics, the calculation of the low-lying
modes of $H_W$ is skipped.
We fix the value of $\lambda_{thrs}$, the lower-bound of the region
where the approximation formula (\ref{eq:Zolotarev}) is valid,
though the modes with $|\lambda| < \lambda_{thrs}$ may appear.
The error with this setting is corrected by the noisy Metropolis test
performed at the end of each MD evolution in addition to the usual
Metropolis test.
This algorithm is twice faster than the case with the 4D solver with
the projection of low-modes of $H_W$, which was used in an early
stage of the simulation.

The $N_f=2+1$ simulation has been started recently
\cite{Poster_Hashimoto}.
The one-flavor part is implemented with one of the chirality
sectors \cite{Bode:1999dd,DeGrand:2006ws} making use of the
fact that $H(m)^2$ commutes with $\gamma_5$, thus
\begin{equation}
  H^2 = P_+ H^2 P_+ + P_- H^2 P_- \equiv Q_+ + Q_-,
\hspace{0.7cm}
 \det H^2 = \det Q_+ \cdot \det Q_- .
\end{equation}
Except for the trivial contribution from the zero-modes,
the determinant of one chirality sector corresponds to
the contribution of one flavor.
Thus, the pseudofermion action
$S_F=\phi_\sigma^\dag Q_\sigma^{-1}\phi_\sigma$,
where $\sigma$ can be either $+$ or $-$, represents the one-flavor
of dynamical fermion.
The same acceleration techniques as $N_f=2$ are applicable to
the one-flavor part.
For the $N_f=2+1$ simulation, we adopt the 5D solver with the projection
of low-modes of $H_W$, which no longer requires the noisy
Metropolis test \cite{Poster_Hashimoto}.

\section{Simulation}
\label{sec:simulation}

Numerical simulations are performed mainly on IBM Blue Gene/L system
at KEK.
At present the sustained performance on one rack of Blue Gene/L
(1024 nodes, 5.7 TFlops of peak performance) 
is about 30\% for the Wilson operator \cite{Talk_Doi},
and 10--15\% for the overlap HMC.

The $N_f=2$ simulations are performed on $16^3\times 32$ lattices
at $\beta=2.3$.
We use 6 quark masses, 0.015, 0.025, 0.035, 0.050, 0.070, and 0.100,
roughly corresponding to $m_s^{phys}/6$ -- $m_s^{phys}$.
10,000 trajectories of a length 0.5 are generated at each sea quark
mass in the $Q=0$ sector.
For $m_{ud}=0.050$ ($\simeq m_s^{phys}/2$), configurations in $Q=-2$ and
$-4$ sectors are also generated.
In addition, we perform a simulation in the $\epsilon$-regime at
$\beta=2.35$ with a quite small quark mass, $m_{ud}\simeq 3$ MeV.
To generate one trajectory, 11--26 minutes are needed on one rack
of Blue Gene/L.
The acceptance rate is kept to 80--90\% at each quark mass.
The locality of the overlap operator was examined
in Ref.~\cite{Yamada:2006fr}.

The $N_f=2+1$ simulations have been started recently on $16^3\times 48$
lattices with almost the same parameters as $N_f=2$,
while each trajectory has a length 1.0 \cite{Poster_Hashimoto}.
We use 5 values of $m_{ud}$ covering the same quark mass region as
the $N_f=2$ case for each of 2 strange quark masses, $m_s=0.080$ and
0.100, around the physical strange quark mass $m_s^{phys}$.
Present performance is around 2 hours for one trajectory
on one rack of Blue Gene/L.

\begin{figure}[tb]
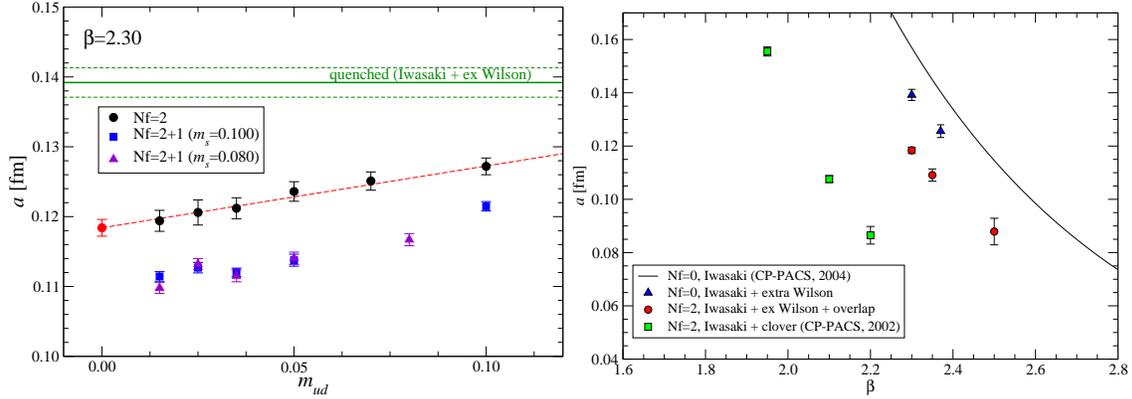

\center{
\includegraphics[clip=true,width=7.4cm]{Figs/a_vs_mq.eps}
\includegraphics[clip=true,width=7.4cm]{Figs/beta_shift.eps}
}
\vspace{-0mm}
\caption{
The left panel shows the lattice scale $a(r_0)$
set by $r_0 = 0.49$fm.
The right panel shows the $\beta$-dependence of $a(r_0)$
for $N_f=2$ and 0, together with the result of the clover fermion.
}
\label{fig:scale}
\vspace{-0mm}
\end{figure}

The lattice scale is set by the hadronic radius $r_0$ which
is defined through
\begin{equation}
 \left.  r^2 \frac{V(r)}{\partial r} \right|_{r=r_0} = 1.65,
\end{equation}
by setting the physical value $r_0=0.49$ fm.
The static quark potential $V(r)$ is calculated with the standard
procedure.
Figure~\ref{fig:scale} shows the result of the lattice spacing.
The left panel displays the result for $N_f=2$ extrapolated to
the chiral limit, as well as a preliminary result for $N_f=2+1$.
A linear extrapolation of the $N_f=2$ data gives
$a(m_{ud}=0)=0.1184(12)_{stat}(11)_{syst}$ fm.
The right panel shows the $\beta$-dependence of $a$ in the chiral
limit for $N_f=0$ and $2$ together with the clover fermion case.
The shift of $\beta(a)$ with respect to the number of
flavors is milder for the overlap fermion than for the Wilson-type
fermions.
This behavior is consistent with the perturbative calculation.

\section{Results}
\label{sec:results}

On the $N_f=2$ lattices, we finished the generation of gauge
configurations and are currently calculating various physical observables.
The following calculations are in progress.
\begin{itemize}
\vspace{-0.12cm}
\item $\epsilon$-regime
\cite{Fukaya:2006xp,Fukaya:2007fb,Fukaya:2007yv,Talk_Fukaya}
\vspace{-0.3cm}
\item Topological Susceptibility \cite{Talk_Chiu,Aoki:2007pw}
\vspace{-0.3cm}
\item Pion mass and decay constant \cite{Noaki:2007es}
\vspace{-0.3cm}
\item Pion form factor \cite{Talk_Kaneko}
\vspace{-0.3cm}
\item $B$ meson bag parameter \cite{Yamada:2007nh}
\vspace{-0.3cm}
\item $\pi^+$-$\pi^0$ mass difference \cite{Shintani:2007ub}
\vspace{-0.3cm}
\item Pion scattering length \cite{Talk_Yagi}
\vspace{-0.12cm}
\end{itemize}
In the following, we briefly describe the first three subjects,
which are the first testing ground of the viability of
our simulations.

\subsection{$\epsilon$-regime}
\label{subsec:epsilon-regime}

The chiral condensate is related to the spectral density of the Dirac
operator through the Banks-Casher relation \cite{Banks:1979yr},
\begin{equation}
 \Sigma \equiv - \langle \bar{q} q \rangle
 = \lim_{m\rightarrow \infty} \lim_{V\rightarrow \infty}
   \frac{\pi \rho(0)}{V} ,
\end{equation}
where $\rho(\lambda) = \sum_k \langle \delta(\lambda-\lambda_k)\rangle$
is the spectral density of the Dirac operator.
The accumulation of low modes generate the spontaneous chiral
symmetry breaking.
Two limits, $m\rightarrow \infty$ and $V\rightarrow \infty$,
in the above equation are not commutable.
However, it is also convenient to consider the opposite order
of the limit.
The $\epsilon$-regime is defined through the condition
\begin{equation}
 1/\Lambda_{QCD} \ll L \ll 1/m_\pi
\end{equation}
which implies $m\ll 1/\Sigma V$.
In the $\epsilon$-regime, the low energy effective theory is
applied with the same parameters as the infinite $V$ case.
Because of the finiteness of the volume, the topological charge
dependence of observables becomes manifest, and hence it is
convenient to determine the parameters of the low energy effective
Lagrangian in the fixed topology simulations.
Another advantage of the $\epsilon$-regime simulations is that
the chiral random matrix theory (RMT) is expected to describe
the behavior of the low-lying modes.

\begin{figure}[tb]
\center{
\includegraphics[clip=true,width=7.8cm]{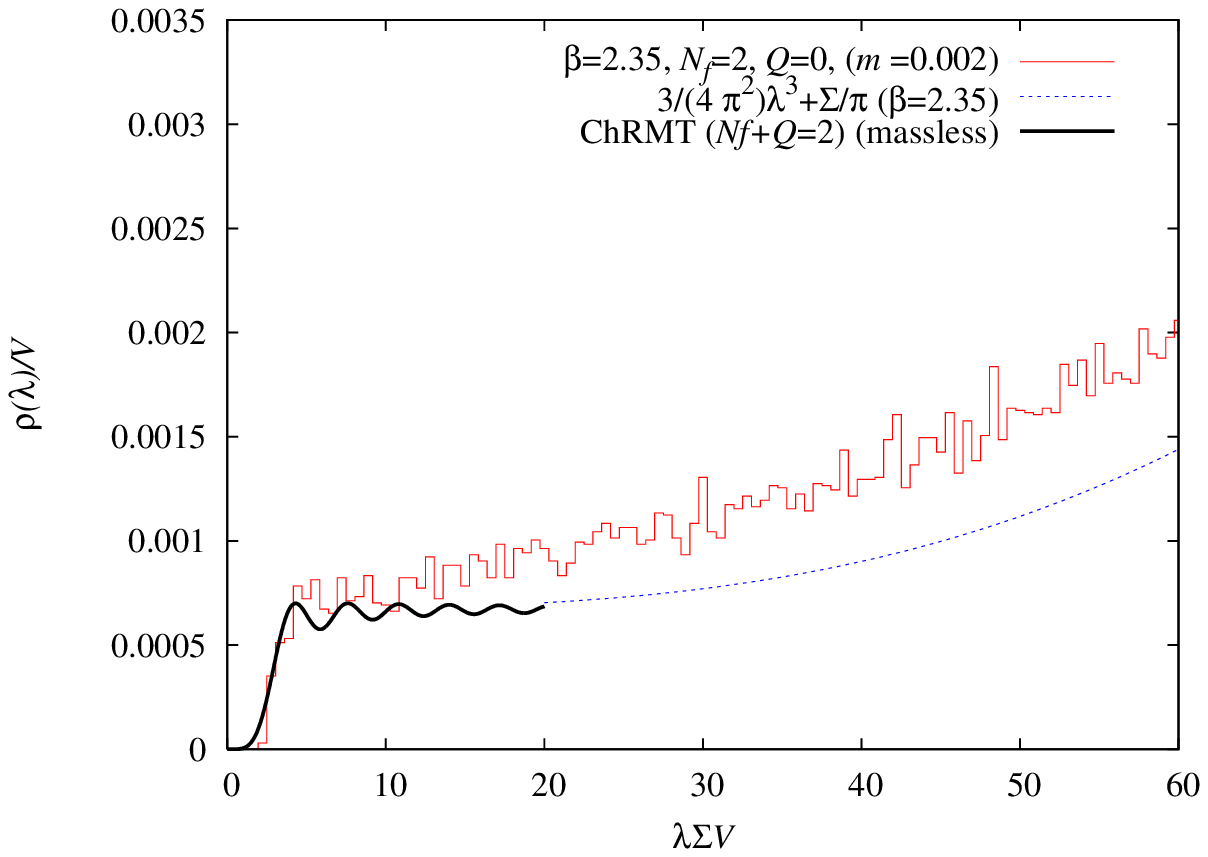}
\hspace{-0.8cm}
\includegraphics[clip=true,width=7.8cm]{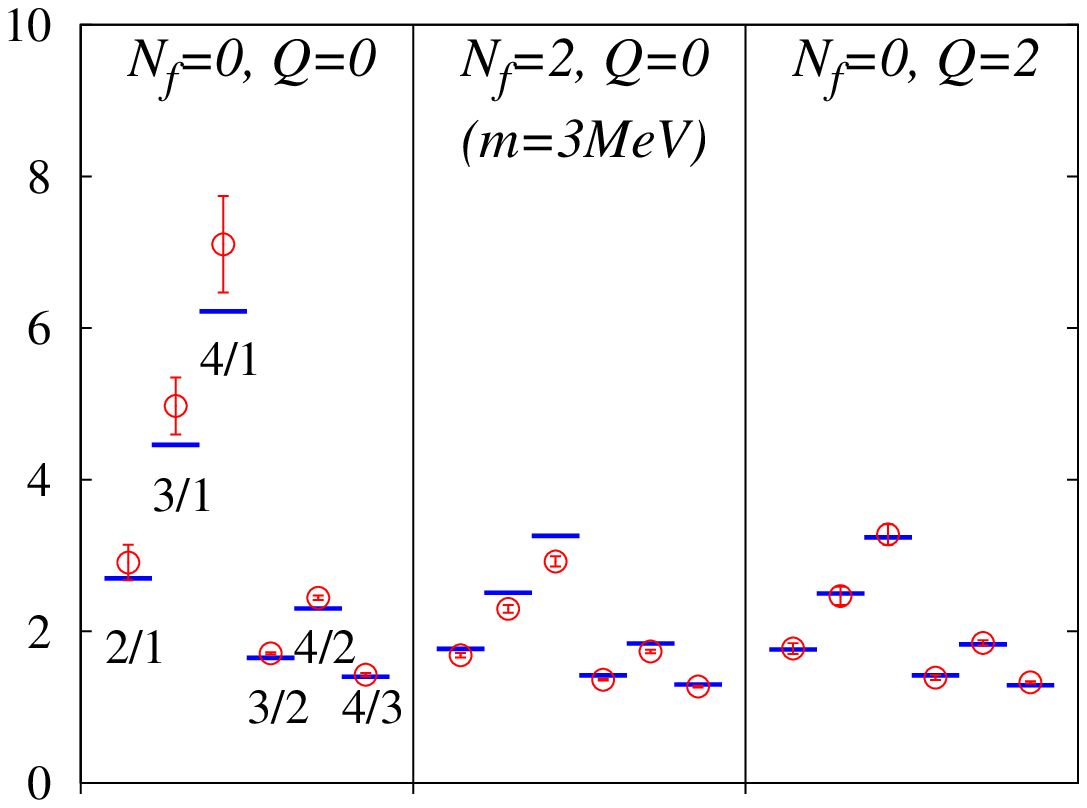}
}
\vspace{-0.6cm}
\caption{
The result in the $\epsilon$-regime at $N_f=2$.
The left panel shows the spectral density of the overlap-Dirac
operator.
The right panel compares the level spacings in the cases of
different $Q$ and $N_f$.
}
\label{fig:e-regime}
\vspace{-0mm}
\end{figure}

The results in the $\epsilon$-regime for $N_f=2$ are displayed in
Figure~\ref{fig:e-regime}
\cite{Fukaya:2007fb,Fukaya:2007yv}.
We find good agreement with the RMT prediction for the
level distribution.
The lowest level distribution gives a value of the chiral
condensate which is consistent with other determinations.
Another agreement with the RMT prediction is found for
the so-called topology-flavor duality, as shown in the right panel
of Figure~\ref{fig:e-regime}.
With the nonperturbative renormalization, we obtaine
\begin{equation}
 \Sigma^{\overline{MS}}(2\mbox{GeV}) =
(\, 251 \pm 7\mbox{\small (stat)} \pm 11\mbox{\small (syst)} \ 
    \mbox{MeV}\,)^3 .
\label{eq:chiral_condensate_e-regime}
\end{equation}
The main source of the systematic error is $O(\epsilon^2)$,
which can be corrected using meson correlators \cite{Talk_Fukaya}.

\subsection{Topological susceptibility}
\label{subsec:topological_susceptibility}

Now let us go back to the $p$-regime.
We first need to show that the topological susceptibility is
successfully determined in the fixed-$Q$ ``vacua'' from the
correlation function (Sec.~\ref{subsec:Simulation_fixed_topology}).
The topological susceptibility $\chi_t$ is extracted from
the correlation function
\begin{equation}
 C_{\eta'}(t) = \frac{1}{L^3} \sum_{\vec{x}}
   \langle m_q P^0(x) m_q P^0(0) \rangle_Q ,
\label{eq:corr_etaprime}
\end{equation}
where $P^{0}(x)$ is the flavor singlet pseudoscalar density.
For the overlap fermions,
\begin{equation}
  P^{0}(x) = \frac{1}{N_f} \sum_{f=1}^{N_f}
   \bar{\psi}^f(x) \gamma_5 \left[ 1 - \frac{aD(m_q=0)}{2m_0}  \right]
   \psi^f(x).
\end{equation}
At large $t$,
\begin{equation}
 C_{\eta'}(t) = \frac{1}{V}\left( \frac{Q^2}{V} - \chi_t
                           - \frac{c_4}{2\chi_t V} \right)
   +  {\cal O}(V^{-3}) + {\cal O}(e^{-m_{\eta'}t}),
\end{equation}
hence the value of $\chi_t$ is extracted from the plateau of
$C_{\eta'}(t)$.

The measurement of $\chi_t$ is performed on $N_f=2$ and $Q=0$
lattices at every 20 trajectories, thus with statistics of 500.
Prior to measuring the correlator, we calculate
50 pairs of low-lying eigenmodes of the overlap operator $D(0)$
by the implicitly restarted Lanczos algorithm.
Since these low-modes can be explicitly inverted, the solver
algorithm is applied to the operator projected out these modes,
whose condition number is reduced from the original one.
This low-mode preconditioning accelerates the calculation
of the quark propagators by a factor of 8.
These low-modes are also used in the low-mode averaging, which
is an average of correlators over all the spacetime source points
for the low-mode contributions
\cite{DeGrand:2004qw,Giusti:2004yp}.
For the disconnected part of the correlator (\ref{eq:corr_etaprime}),
the quark propagator is approximated by 50 pairs of the 
low-eigenmodes, by observing these modes dominate the correlator.

\begin{figure}[tb]
\center{
\includegraphics[clip=true,width=6.8cm]{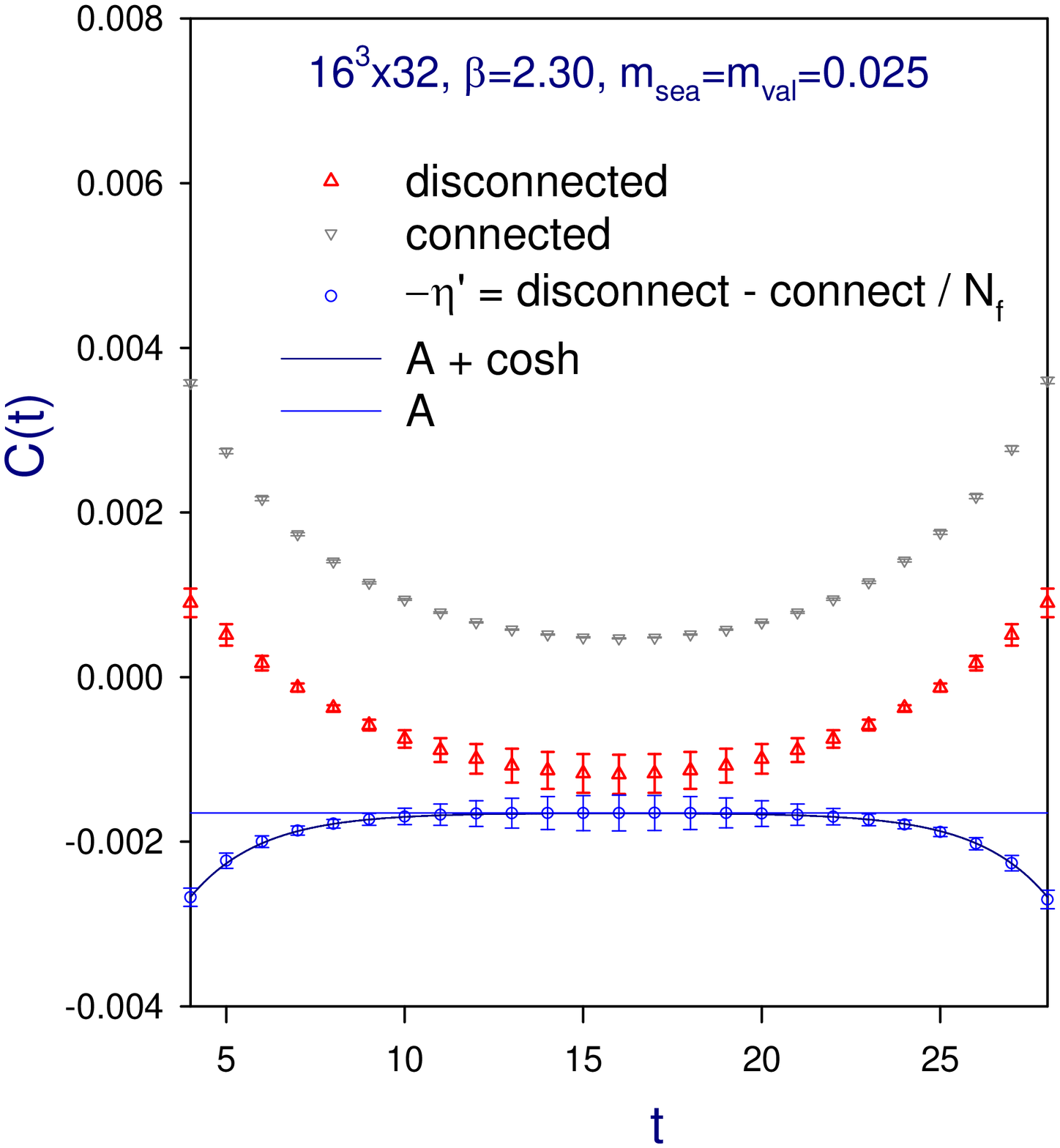}
\includegraphics[clip=true,width=7.8cm]{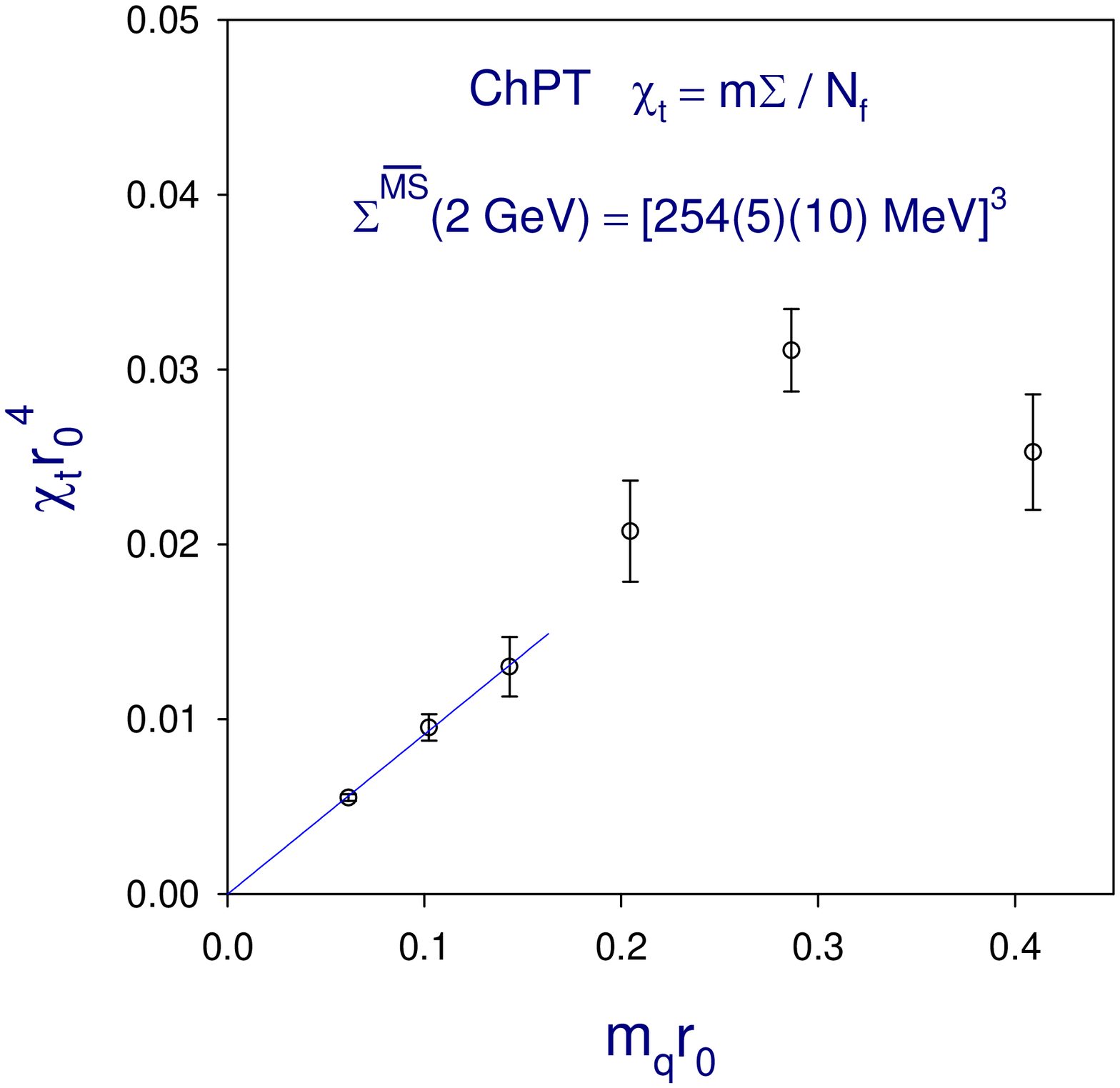}
}
\vspace{-2.mm}
\caption{
The result of the topological susceptibility for the $N_f=2$ simulation.
The left panel shows the extraction of the topological susceptibility
$\chi_t$ from the $\eta'$ correlator at $m_q=0.025$.
The right panel displays $\chi_t r_0^4$ versus the sea quark mass
$m_q r_0$.
}
\label{fig:top_suscept}
\vspace{-0mm}
\end{figure}

The result is shown in Figure~\ref{fig:top_suscept}.
The left panel displays the correlator $-C_{\eta'}(t)$ at
$m_q=0.025$.
The data are fitted to a function $A+B(e^{-Mt} + e^{-M(T-t)})$.
Assuming $|c_4|\ll 2\chi_t V$, we obtain
$a^4\chi_t = 3.40(27)\times 10^{-5}$ at $m_q=0.025$.
The right panel shows the extracted $\chi_t$ as a function of
sea quark mass in units of $r_0$.
For the smallest three quark masses, the data are well fitted to
a linear function whose intercept is consistent with zero.
This behavior is consistent with the prediction of the chiral
perturbation theory \cite{Leutwyler:1992yt},
\begin{equation}
 \chi_t = \frac{m_q \Sigma}{N_f} + {\cal}O(m_q^2).
\end{equation}
From the slope of the fit result, a value of the chiral condensate
is obtained as $r_0^3 \Sigma=0.182(6)$.
In order to convert it to a physical value, we use the renormalization
factor $Z_m^{\overline{\rm MS}} (2 \mbox{GeV})=0.742(12)$, which is
obtained nonperturbatively through the RI/MOM scheme on the
lattice \cite{Martinelli:1994ty,Noaki:2007es}.
This leads to a value
\begin{equation}
 \Sigma^{\overline{MS}}(2\mbox{GeV}) =
(\, 254 \pm 5\mbox{\small (stat)} \pm 10\mbox{\small (syst)}\ 
    \mbox{MeV}\,)^3 ,
\label{eq:chiral_condensate_topsus}
\end{equation}
which is consistent with the value in
Eq.~(\ref{eq:chiral_condensate_e-regime}) independently obtained 
in the $\epsilon$-regime.
The statistical error includes those of $a^{-1}$ and
$Z_m^{\overline{\rm MS}}$, and the systematic error is of the higher
order effects such as the $c_4$ term.

We also measure $\chi_t$ on the $N_f=2$ lattices with $Q=-2$ and $-4$
at $m_q=0.050$ with statistics of 250 configurations.
The extracted values of $\chi_t$ are consistent with the value
at $Q=0$.

These results indicate that the local fluctuation of the topological
charge is active enough  to produce relevant chiral dynamics
on the fixed-$Q$ vacua.
The successful extraction of $\chi_t$ enables us to quantify
the finite size effects for other observables
represented in Eq.~(\ref{eq:Corr_fixedQ}).

\subsection{Pion mass and decay constant}

Among hadronic observables, the pion mass and decay constant are
the primary quantities to be tested with the chiral perturbation theory.
Here we briefly present our results of the $N_f=2$
simulation for $m_\pi$ and $f_\pi$ \cite{Noaki:2007es}.

The meson correlators are computed at every 20 trajectories.
The low-mode preconditioning and the low-mode averaging
described in the previous subsection are applied to the meson
correlators.
The pion mass and decay constant are extracted from
the pseudoscalar meson correlators with point and smeared sources
by a simultaneous fit.
$f_\pi$ is obtained through the axial Ward-Takahashi identity,
\begin{equation}
  f_\pi = 2m_q \langle 0 | P(0)|\pi \rangle/m_{\pi}^2 ,
\end{equation}
without further renormalization.
Note that we are using the $f_\pi=130$ MeV normalization.
The quark mass is renormalized with the renormalization factor
$Z_p^{\overline{\rm MS}} (2 \mbox{GeV})=0.742(12)$, which is
obtained nonperturbatively through the RI/MOM scheme on the
lattice \cite{Martinelli:1994ty,Noaki:2007es}.

The pion mass and decay constant receive two kinds of finite size
effect (FSE): the standard FSE and that from fixed $Q$.
The former effect is evaluated by the formulae determined by
Colangelo {\it et al.} \cite{Colangelo:2005gd},
which is developed from the L\"uscher's formula for the relation
between the pion scattering amplitude and the mass-shift
in a finite box \cite{Luscher:1985dn}.
We use the NNLO results of Ref.~\cite{Colangelo:2005gd} and
the values of low energy constants estimated
in Ref.~\cite{Colangelo:2001df}.
For our lightest quark mass, they amount to 4.5\% and 6.0\%
for $m_\pi^2/m_q$ and $f_\pi$, respectively.
The FSE from the fixed topological charge is subject to
Eq.~(\ref{eq:Corr_fixedQ}).
The $O(V^{-1})$ corrections to $m_\pi$ and $f_\pi$ are determined
at NLO of ChPT.
The correction to $m_\pi$ starts at the tree-level.
In the small quark mass region, it shifts the value of $m_\pi^2/m_q$
with the same size as the standard FSE while in the opposite direction.
Thus the two effects almost cancel each other.
The fixed-$Q$ correction to $f_\pi$ starts at NLO of ChPT, and
gives small effect on the result of $f_\pi$.

\begin{figure}[tb]
\center{
\includegraphics[clip=true,width=7.3cm]{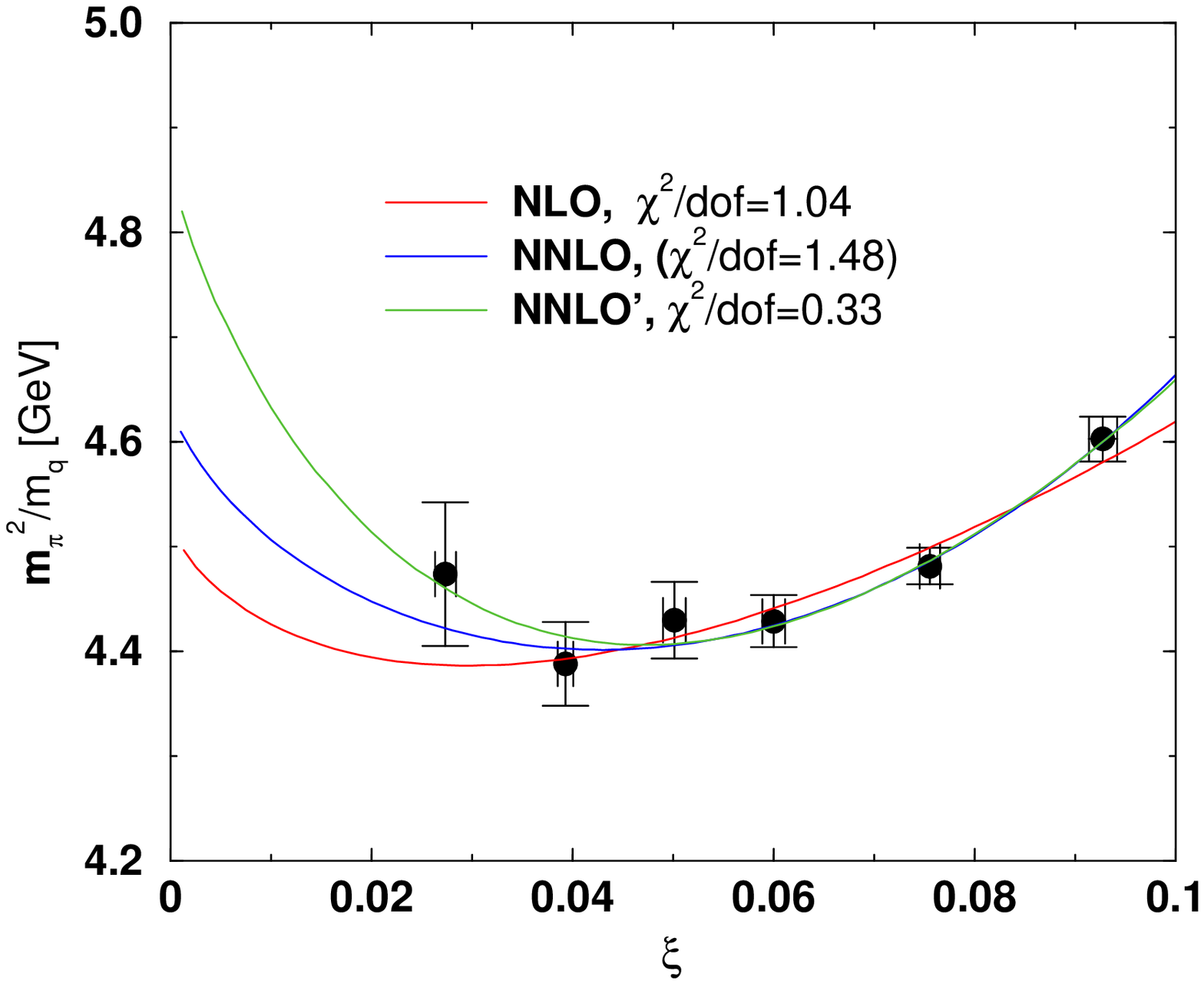}
\includegraphics[clip=true,width=7.5cm]{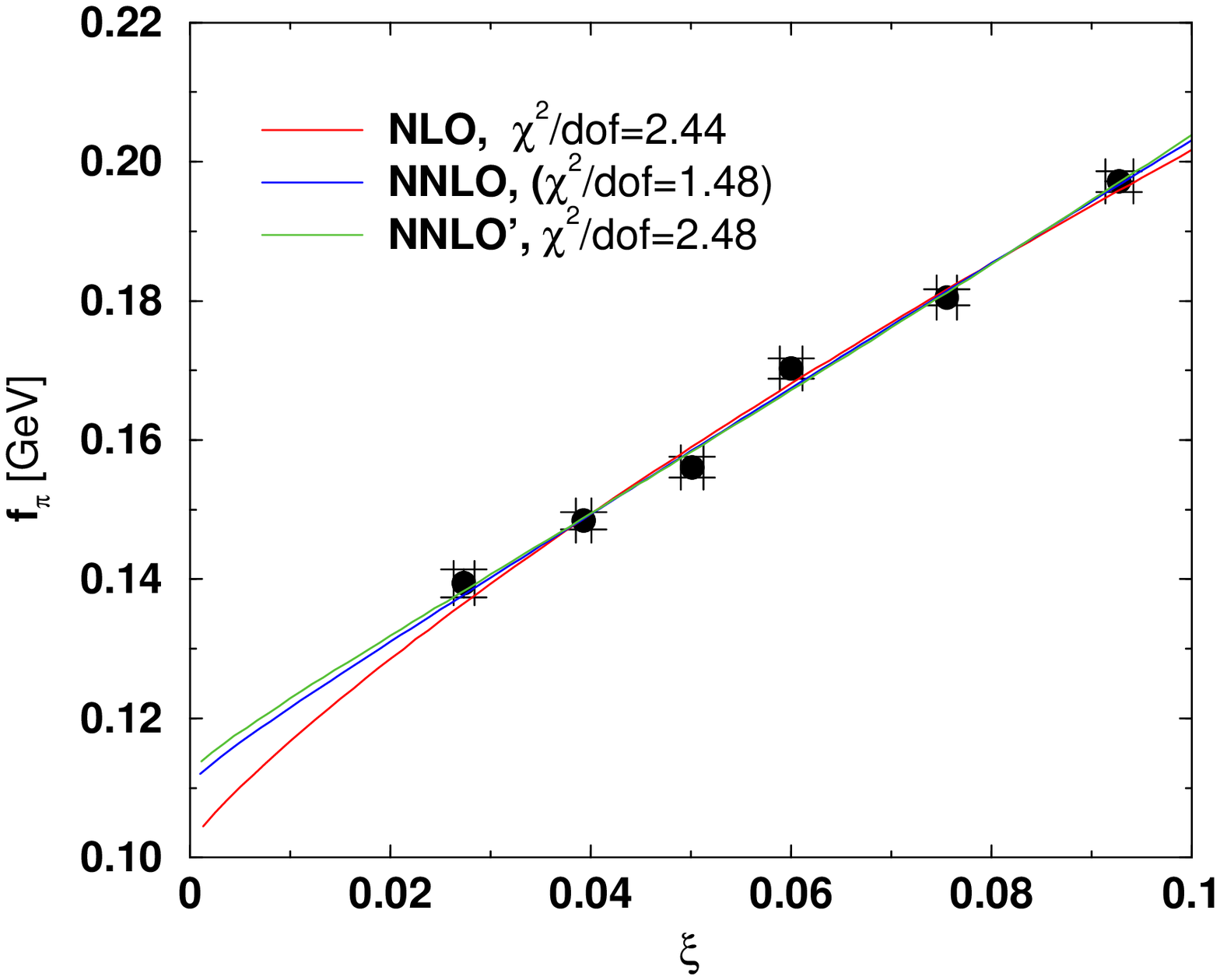}
}
\vspace{-0.2cm}
\caption{
The results of chiral extrapolation for $m_\pi^2/m_q$ (left panel)
and $f_\pi$ (right) in physical units as functions of $\xi$.
Three types of fits, NLO, NNLO, and NNLO' are performed using
all the available data points.
}
\label{fig:pion_mass_decayconst}
\vspace{-0mm}
\end{figure}

\begin{figure}
 \begin{center}
  \hspace{0.5mm}
  \includegraphics[width=6.5cm,clip]{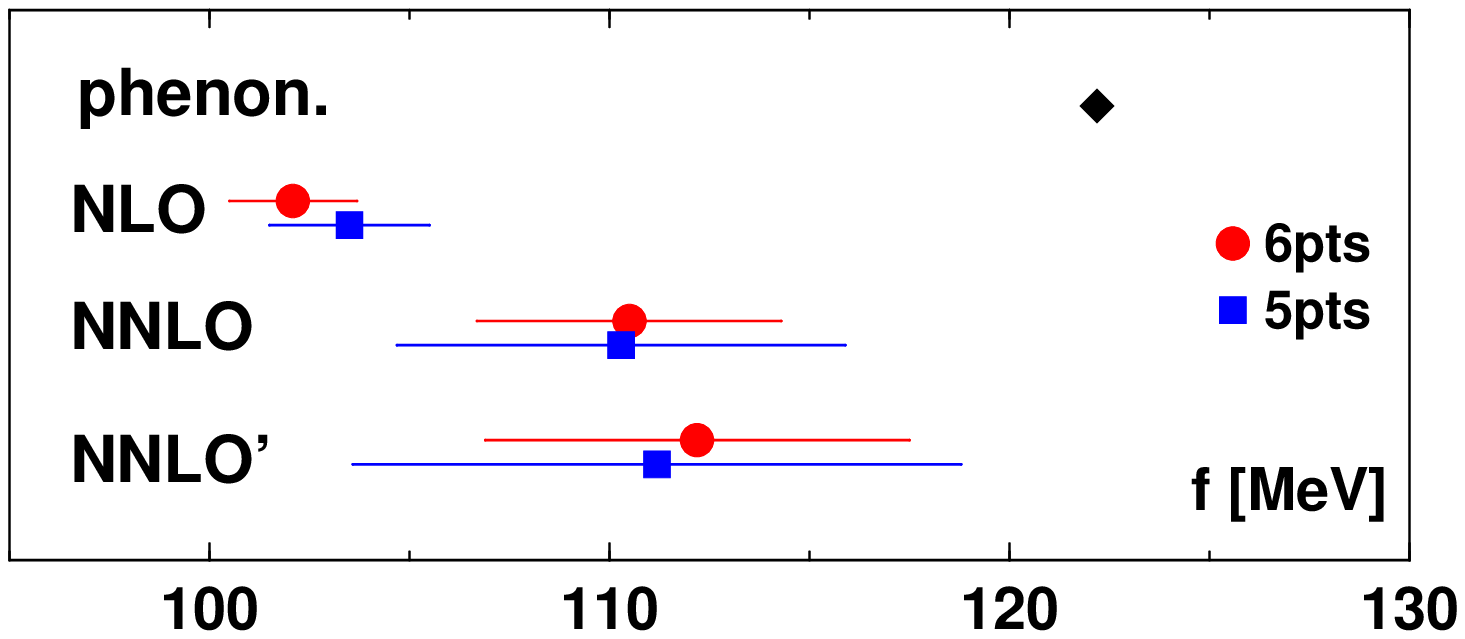} 
  \hspace{0.2mm}
  \includegraphics[width=6.5cm,clip]{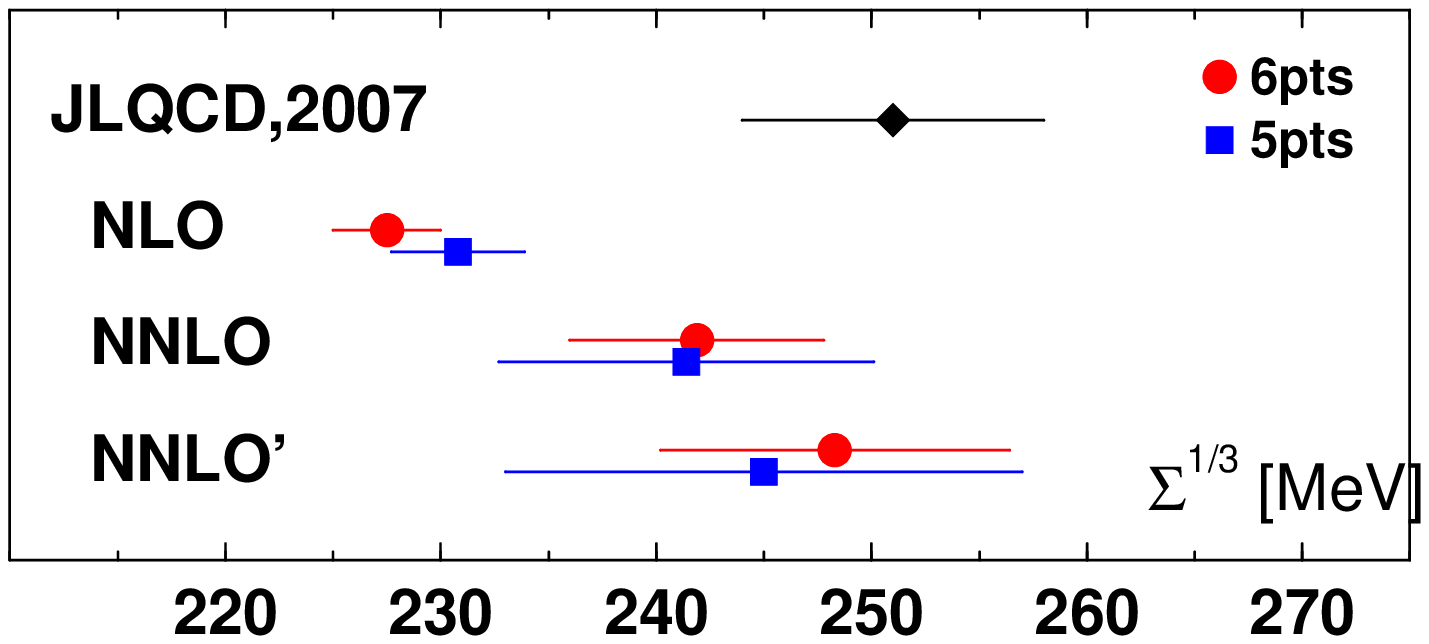}
 \vspace{0.3cm} \\
  \includegraphics[width=6.4cm,clip]{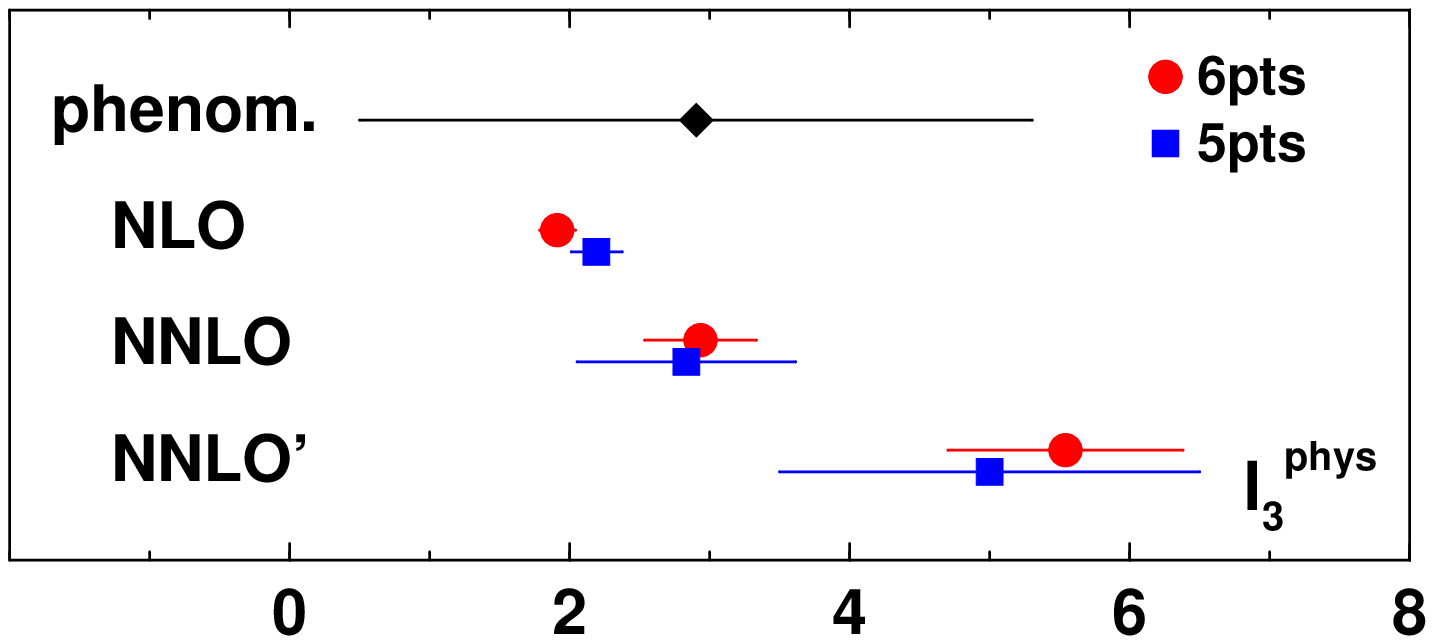} 
  \hspace{1mm}
  \includegraphics[width=6.4cm,clip]{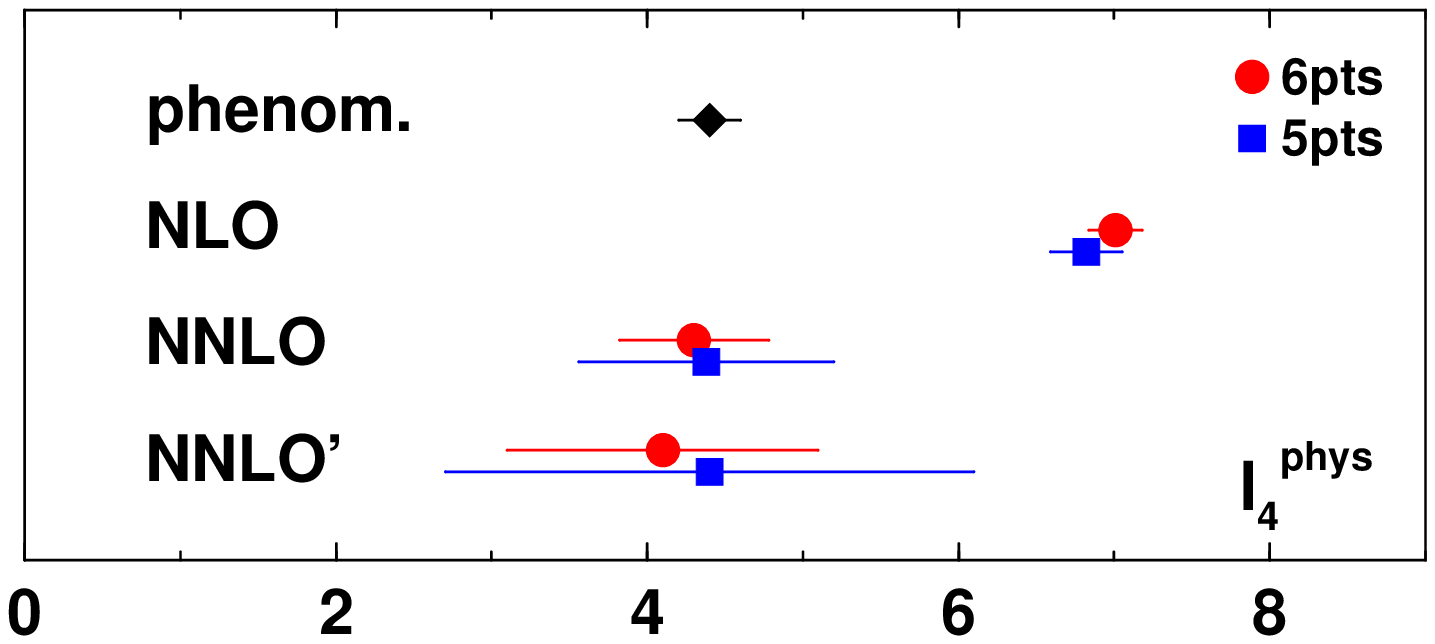} 
  \end{center}
\vspace{-0.4cm}
\caption{
Comparisons of physical quantities obtained from the
chiral fit ansatze: $f$ (upper left), $\Sigma^{1/3}$ (upper right), 
$\bar{l}_3^{\rm phys}$ (lower left) and  
$\bar{l}_4^{\rm phys}$ (lower right).
In each panel, the red circles and 
blue squares are  corresponding to the fits with 6 and 5 lightest
data points, respectively.}
\label{fig:low_energy_consts}
\end{figure}

Figure~\ref{fig:pion_mass_decayconst} shows the results of
$m_\pi^2/m_q$ and $f_\pi$ after correcting FSEs.
To compare with the ChPT expansion, we chose
$\xi\equiv (m_\pi /4\pi f_\pi)^2$ as an expansion parameter,
where $f_\pi$ is the measured value at each quark mass.
We apply three types of ChPT fit: NLO, NNLO, and 
a simplified NNLO called NNLO' hereafter.
The NNLO' formulae are obtained from those of NNLO by
using an approximation $\xi^2\ln \xi\approx -2.5 \xi^2$,
which is numerically reasonable in our target range,
$0< \xi \les 0.1$.
Explicit functional forms are given in Ref.~\cite{Noaki:2007es}.
The results of fits are displayed in
Figure~\ref{fig:pion_mass_decayconst}.
We note that the fit to the NNLO formulae is performed by a simultaneous
fit for $m_\pi^2/m_q$ and $f_\pi$, and thus the value of $\chi^2$
cannot be compared directly to other fits.

These fits determine the following quantities: $f$,\
$\Sigma =B_0\cdot f^2/2$,\ $\bar{l}_3^{\rm phys}$, 
and $\bar{l}_4^{\rm phys}$.
Figure~\ref{fig:low_energy_consts} compares the results of fits
with the phenomenologically determined values 
($f$ \cite{Gasser:1983yg} and $\bar{l}_3^{\rm phys}$
$\bar{l}_4^{\rm phys}$ \cite{Colangelo:2001df}).
The result of $\Sigma$ is compared with our calculation
in the $\epsilon$-regime in Sec.~\ref{subsec:epsilon-regime}
\cite{Fukaya:2007fb}.
Except for the case of $\bar{l}_3^{\rm phys}$,
the results of NLO fit are inconsistent with NNLO and the phenomenological
estimates.
This implies the failure of the NLO formulae to describe our data.
The results of NNLO and NNLO' fits are consistent with each other
and with the phenomenological estimates.

To quantify the FSEs more accurately, we need simulations with
a larger lattice size, as well as a comparison with the result at
$Q\neq 0$.
We emphasize that these results indicate that the overlap simulations
with fixed topology provide a framework for precision calculations
of the spectrum and the matrix elements in the chiral regime.

\section{Conclusion}

We are performing large-scale simulations with $N_f=2$
and $N_f=2+1$ dynamical overlap fermions.
The range of sea quark mass covers $m_s^{phys}/6$ -- $m_s^{phys}$.
The simulations are performed in fixed topological charge sectors.
The results of the topological susceptibility and the
pion mass and decay constant indicate that these simulations
can provide a ground for precision computations of matrix elements
with controlled chiral extrapolation.

We completed a generation of $N_f=2$ gauge configurations on
$16^3\times 32$ lattices with $a\simeq 0.12$ fm.
Numbers of measurements are in progress and planned.
The $N_f=2+1$ simulations on $16^3\times 48$ lattices are also
in progress.
These configurations will be supplied to ILDG soon after
the first publication of the result of the spectrum.

For further investigation of finite size effects and for
more extended objects than mesons, simulations at
larger lattice sizes are desired.
The target size of the spatial extent is 24, which requires
further improvements of numerical algorithms.

\section*{Acknowledgment}

Numerical simulations were performed on IBM System Blue Gene 
Solution and Hitachi SR11000 at High Energy Accelerator Research
Organization (KEK) under a support of its Large Scale Simulation
Program (No. 06-13, 07-16).
We thank J.~Doi, H.~Samukawa, and S.~Shimizu of IBM Japan
Tokyo Research Laboratory for tuning the QCD
code on the Blue Gene.
The numerical simulations were partly performed on NEC SX8 at
Yukawa Institute for Theoretical Physics, Kyoto University,
and on NEC SX8 at Research Center for Nuclear Physics,
Osaka University.
The simulation also owes to a gigabit network SINET3 supported by
National Institute of Informatics, for efficient data transfer
supported by JLDG.
This work is supported in part by the Grant-in-Aid of the
Japanese Ministry of Education (No.~19740160).

\end{document}